\documentclass{article} 
\usepackage{iclr2021_conference,times}

\usepackage{amsmath,amsfonts,bm}

\def\eqref#1{equation~\ref{#1}}

\def\1{\bm{1}}

\DeclareMathAlphabet{\mathsfit}{\encodingdefault}{\sfdefault}{m}{sl}
\SetMathAlphabet{\mathsfit}{bold}{\encodingdefault}{\sfdefault}{bx}{n}

\usepackage{hyperref}
\usepackage{url}
\usepackage{soul}
\usepackage{algorithm}
\usepackage{algpseudocode}
\usepackage[font=scriptsize,labelfont=bf]{caption}

\usepackage{graphicx}

\title{Modally Reduced Representation Learning of Multi-Lead ECG Signals through Simultaneous Alignment and Reconstruction}

\author{Nabil Ibtehaz \thanks{ Work done as a summer intern at Futurewei Technologies.} \\
Department of Computer Science\\
Purdue University\\
West Lafayette, IN, USA \\
\texttt{nibtehaz@purdue.edu} \\
\And
Masood Mortazavi \\
IC Lab \\
Futurewei Technologies Inc.\\ Santa Clara, CA, USA \\
\texttt{masood.mortazavi@futurewei.com}
}

\iclrfinalcopy 
\begin{document}

\maketitle

\begin{abstract}

Electrocardiogram (ECG) signals, profiling the electrical activities of the heart, are used for a plethora of diagnostic applications. However, ECG systems require multiple leads or channels of signals to capture the complete view of the cardiac system, which limits their application in smartwatches and wearables. In this work, we propose a modally reduced representation learning method for ECG signals that is capable of generating channel-agnostic, unified representations for ECG signals. Through joint optimization of reconstruction and alignment, we ensure that the embeddings of the different channels contain an amalgamation of the overall information across channels while also retaining their specific information. On an independent test dataset, we generated highly correlated channel embeddings from different ECG channels, leading to a moderate approximation of the 12-lead signals from a single-channel embedding. Our generated embeddings can work as competent features for ECG signals for downstream tasks.

\end{abstract}

\section{Introduction}

Electrocardiogram (ECG) signals are used in the diagnosis of cardiovascular diseases---the leading cause of death worldwide, claiming a life every 33 seconds in the USA \footnote{National Center for Health Statistics. Multiple Cause of Death 2018–2021 on CDC WONDER Database. Accessed February 2, 2023.}. ECG signals represent the electrical activities of a heart during the cardiac cycle \cite{Hannun2019Cardiologist-levelNetwork}. Utilizing ECG signals to their fullest potential requires conventional 12-lead ECG recorder~\cite{H.Bunce202030.Cardiology}. This limits the use cases of near-clinical ECG in wearables and smartwatches. Hence, the growing interest in developing competent diagnosis algorithms with a limited set of channels\cite{Reyna2021Will2021}, which introduces new levels of complexity and algorithmic challenges \cite{Reyna2022IssuesPopulations}.

Oriented to specific applications, ECG investigations have used a variety of feature extraction methods, e.g., transformations \cite{Mohonta2022ElectrocardiogramModel}, decompositions \cite{Ibtehaz2019VFPred:Signals}, fiducial point features \cite{Ullah2023DCDA-Net:Electrocardiograms}, etc. However, features competent for one application seldom translates well to a different one \cite{Singh2023ECGApplications}. Observing the impact of representation learning in a multitude of domains \cite{Bengio2013RepresentationPerspectives}, it becomes desirable to develop general representation learners for ECG \cite{Mehari2022Self-supervisedData}, most notably using masked autoencoders \cite{He2021MaskedLearners}, which have proven to be quite competent~\cite{Sawano2022MaskedDysfunction., Yang2022MaskedLearning, Zhang2023MaeFE:Learning, Wang2023UnsupervisedAnalysis}. However, such models were assessed on a specific task \cite{Sarkar2022Self-SupervisedRecognition} and sometimes evaluated on only a portion of the pretraining dataset \cite{Sawano2022MaskedDysfunction.} while requiring all 12 channels \cite{Zhang2023MaeFE:Learning} for satisfactory performance.

While these approaches partially address the issue of task-specific feature extraction, the requirement for multiple ECG leads persists. 
To explore mode (ECG lead) reduction, one can adopt the perspective of multi-modal learning.
Meta-learning has been used to reconstruct missing modes from available ones~\cite{Ma2021SMIL:Modality}, and other solutions have relied on a shared embedding to achieve this~\cite{Wang2023Multi-modalModelling}. 
Semantic alignment of heterogeneous data modes in a shared vector space has also been used through joint learning of co-embeddings~\cite{Mortazavi2020Speech-ImageTasks}. ECG signals represent a special case of homogeneous multi-modal data, i.e., cardiac electric activity of the same heart from different viewpoints (leads). As a result, some correlations already exist between the different modalities, i.e., channels of the ECG signal; for instance, channel II can be treated as the sum of channels I and III based on Einthoven's law \cite{Silverman1992WillemElectrocardiography}. 

In this work, we propose a unified ECG representation learning method that is channel-independent based on simultaneous alignment and reconstruction.
We achieve this by representing the different ECG channels in a joint embedding space, maintaining semantic alignment through contrastive learning while at the same time using an auto-encoder-based reconstruction objective so that the essential channel-specific information can survive in the alignment process.

\section{Method}
\textbf{\textit{Abstractions and High-level Assumptions.}}
As activity snapshots, the 12-lead ECG signal  $x(t)=\{x^{[1]}(t),x^{[2]}(t),x^{[3]}(t),\dots, x^{[12]}(t)\} \in \mathbb{R}^{12}$ represent the state of a heart $\mathcal{H}$ at an instant $t$, where each channel $x^{[i]}(t)$ corresponds to a projection $P_i$, i.e., $x^{[i]}(t) = P_i(\mathcal{H}_{@t})$. 
We train encoder networks $\mathcal{E}_i, i=1\dots12$ to extract  latent representations $h^{[i]}(t)$ of a heart from the input signals $x^{[i]}(t)$. 
The representations of the different channel signals of the \textit{same} heart should be correlated, i.e., $h^{[i]}(t) \equiv \mathcal{E}_i(x^{[i]}(t)) \approx \mathcal{H}_{@t}, i=1\dots12$ and should also be ``complete'' in the sense that individual signals can be retrieved from them by a set of decoders, i.e., $\mathcal{D}_i, i=1\dots12$, i.e., $\hat{x}^{[i]}(t) \equiv \mathcal{D}_i(h^{[i]}(t))  \approx x^{[i]}(t)$.

\textbf{\textit{Core Model Architecture.}} For our representation learner ($\mathcal{E}_i-\mathcal{D}_i$), we have developed a 1-D version of the original Masked AutoEncoder (MAE) architecture~\cite{He2021MaskedLearners}. Our architecture follows the conventional style of having a strong encoder in conjunction with a weak decoder, and vanilla ViT layers \cite{Dosovitskiy2021AnScale} are used to as the building blocks. Our encoder and decoder have 12 and 8 blocks, with embedding dimension of 768 and 512, respectively. The model takes $T=5\text{ sec}, f_s=500\text{ Hz}$ signals as input and breaks them into $100\text{ samples or } 0.2 \text{ sec}$ long non-overlapping patches, for further processing.


\textbf{\textit{Alignment and Reconstruction Losses.}} We distribute the task of 12-lead ECG representation learning to 12 1-D MAE models each assigned to a particular channel. Our models encode-decode to reconstruct their designated channels while producing correlated embeddings as approximation of the latent heart state by jointly optimizing on alignment and reconstruction losses. Expected mean-squared error (MSE) of the predictions $\hat{x}^{[i]}(t)$ over a minibatch $B$ provides a reconstruction loss:
\begin{equation}
    \text{reconstruction loss} = \mathbb{E}_B [ \frac{1}{12 \times T}\sum_{i=1}^{12}\sum_{t=0}^{T}(x^{[i]}(t)-\hat{x}^{[i]}(t))^2 ]
\end{equation}
Triplet loss helps correlate the $h^{[i]}(t), i=1\dots12$ of different channel signals of the same heart:
\begin{equation}
    \text{alignment loss} = {Triplet}_{B}(A, P, N)
\end{equation}
Here, anchor $A$ corresponds to all the signals in the minibatch, positive $P$ and negative $N$ correspond to signals of different channels from the same vs. different hearts, respectively. 

\textbf{\textit{Combining Losses.}} Jointly optimizing these losses often led to trivial embeddings---highly correlated but unable to reconstruct the signals. To prevent this, we designed a curriculum which gradually shifts focus from reconstruction to alignment losses:
\begin{equation}
    Loss_{@epoch-i} = sin(\frac{i}{N_{epochs}}\frac{\pi}{2})\times \text{alignment loss} + cos(\frac{i}{N_{epochs}}\frac{\pi}{2})\times \text{reconstruction loss}
\end{equation}

\textbf{\textit{Pretraining Dataset.}}
As a large 12-lead ECG signal database for pretraining, we selected the PhysioNet Computing in Cardiology Challenge 2021 dataset \cite{Reyna2021Will2021,PerezAlday2020Classification2020,Goldberger2000PhysioBankPhysioNet}, which includes 88,000 public ECG recordings compiled from 6 sources. We only used ECG signals recorded at 500 Hz sampling frequency ($f_s$) to train our models. The recordings from PTB and INCART databases were used as test dataset. For training and inference, a random $T_s = 5$ s long window was cropped from the signals, and mean normalization was performed without any additional filtering or preprocessing steps. 

\textbf{\textit{Distributed Training.}} We adopted a distributed training protocol where MAE models are trained separately on their individual channels, with a coordinator exchanging embeddings required for computing alignment losses. We used AdamW \cite{Loshchilov2019DecoupledRegularization} with default parameters to jointly optimize reconstruction and alignment.
The models were trained with a base leaning rate of $1e^{-3}$ accompanied by a cosine learning rate scheduler, for 200 epochs with a batch size of 256.

\section{Results}

We assessed different aspects of our method and some downstream tasks on several independent test datasets varying in terms of acquisition protocol, demographics, signal quality, sampling rate, etc.

\textbf{\textit{Correlating Embeddings of Dissimilar Signal Channels.}}
Our primary objective is to develop a unified representation for ECG signals irrespective of the signal channel. 
In order to validate our joint optimization strategy, we took 590 signals from PTB \cite{R.Bousseljot1995NutzungInternet.} and INCART \cite{Tihonenko2008StDatabase} databases, with sampling rates of 1000 and 257 Hz, respectively. 
For randomly cropped 5-second-long signals, we computed the similarity between signals of two channels and the corresponding embeddings using cross-correlation coefficient and cosine similarity, respectively. We performed this random experiment 10 times for each channel pair. The overall correlation, presented in Fig. 1B, reveals that our pipeline managed to generate strongly correlated embeddings, even when the signals are weakly correlated (e.g. V5-aVR). Conversely, for already correlated channel pairs, our embeddings either retained (e.g. V1-V2) or improved (e.g. V4-V5) the correlations.

\textbf{\textit{Reconstructing Disparate Signals from Unified Embeddings.}}
In order to assess our reconstruction performance, we have used 827 ECG recordings used in \cite{Ribeiro2020AutomaticNetwork} ($f_s =400 $Hz). In Fig. 1C, we have presented an example of $2 $ sec reconstruction of a random sample. Since, our generated embeddings are correlated, we have tried to reconstruct all the channel signals both from the native channel and channel 1 signals individually, after masking $75\%$ of the signals. Overall, the reconstructed signals follow the morphology of the original signal, the model failed to reach the full amplitude of the R-peaks, despite predicting a sharp peak. This is a limitation of MAE, which usually reconstructs blurry images \cite{He2021MaskedLearners}, and for ECG reconstruction, reduces the signal magnitude to a fraction \cite{Sawano2022MaskedDysfunction.}. Additionally, we computed mean absolute error values of reconstructing each of the channel signals from the different input channels (Fig. 1D), and we can observe that due to the correlation of our generated embeddings, the reconstruction losses stay within $\pm 5\%$ of the loss for native channel as input. Interestingly, for V1, the native embeddings seem to perform worse, which is also apparent from Fig. 1C. This is likely due to the V1 channel signals of this dataset being substantially different from our pretraining dataset.

\textbf{\textit{Improving Single-Channel Myocardial Infraction Prediction.}}
The diagnosis of myocardial infraction, one important application of ECG signals, substantially benefits from multi-channel information \cite{Gupta2021DeepElectrocardiograms}. Therefore, we aim to investigate whether the proposed unification of the embeddings across channels can help improve single-channel predictions.
In order to assess this, we trained baseline MAE models for the individual channels without any embedding alignment and performed a 5-fold cross-validation on the PTB dataset \cite{R.Bousseljot1995NutzungInternet.}. Next, for each of the different channels, embeddings were extracted from the correlated and baseline MAE models and logistic regression classifiers were used. From the F1 scores presented in Fig. 1F, it can be observed that the correlated models performed better in all the cases. They not only managed to improve the less informative channels (I and aVL) but also retained the superior performance of the highly informative channels (V5 and V6)~\cite{Gupta2021DeepElectrocardiograms}. Therefore, our proposed embedding correlation injects more information to less informative channels, without hampering the representations of the highly informative channels.

\textbf{\textit{Implicitly Learning ECG Biometric Authentication.}}
Since the alignment objective distinguishes and aligns ECG embeddings from an individual's heart, we used the ECG-ID benchmark dataset~\cite{Lugovaya2005BiometricElectrocardiogram, Nemirko2005BiometricElectrocardiogram} to assess model performance in biometric authentication. In order to capture the entire ECG signature, we avoided any masking during this inference and the class token embedding was obtained from the model. A t-SNE~\cite{vanderMaaten2008VisualizingT-SNE} plot presented in Fig 1E shows that the ECG embeddings of different individuals are clustered together. Moreover, as a true assessment of the quality of our embeddings, we used a 1-KNN classifier and achieved an accuracy of $99.68\%$ from 10-fold cross-validation. This is indeed impressive as without any finetuning, on an independent test dataset, we are on par with state-of-the-art methods~\cite{Ibtehaz2022EDITHAuthentication} which are explicitly trained on that dataset for this task.

\begin{figure}
    \centering
    \includegraphics[width=\textwidth]{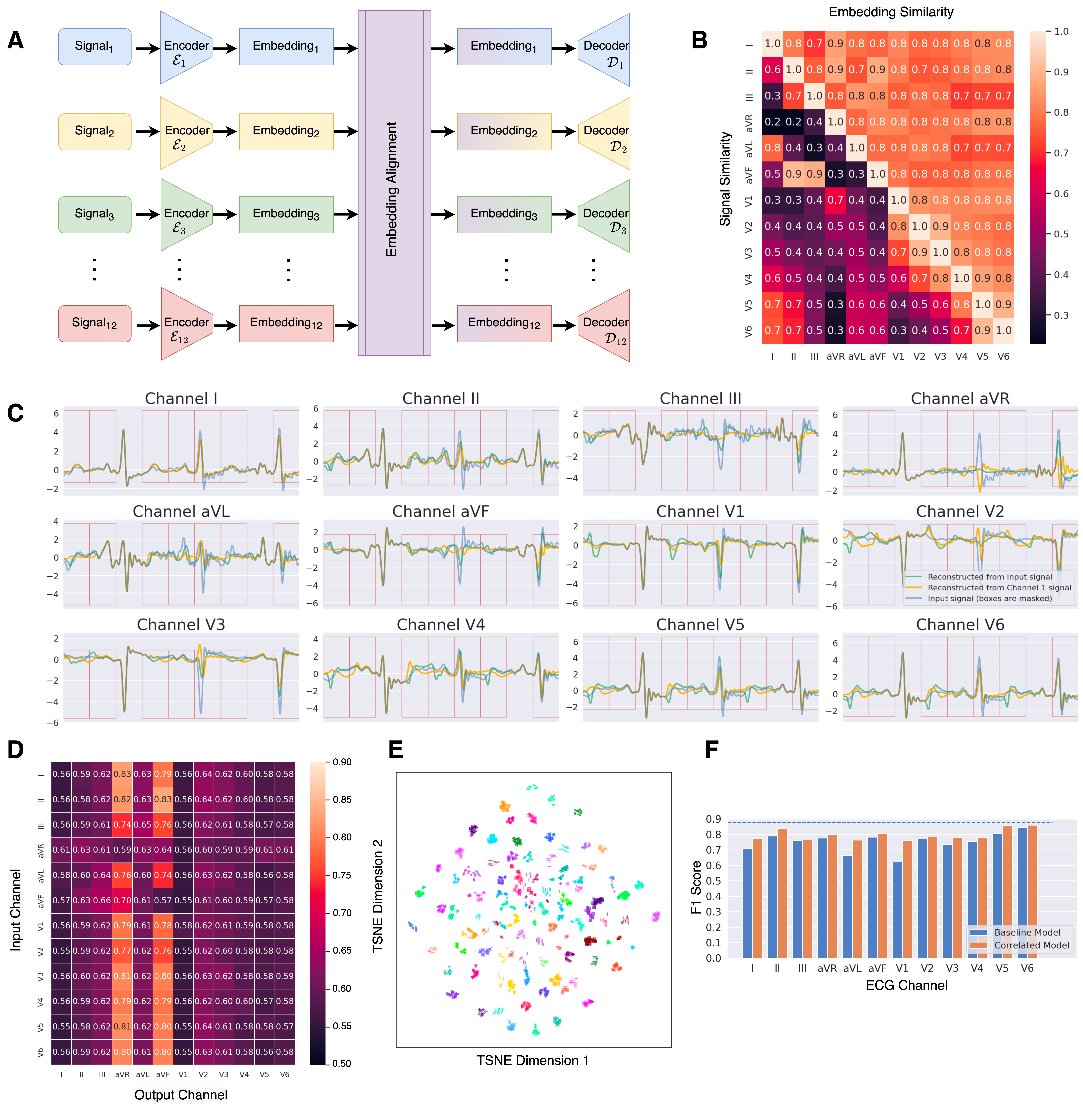}
    \caption{A) Our proposed architecture for correlating Masked AutoEncoders. (B) Signal (lower triangle) versus embedding similarity (upper triangle) for pairs of ECG channels on INCART and PTB datasets. (C) 2s ECG reconstruction of a sample from the test set: blue represents the original signal, while green and yellow correspond to reconstruction from the native versus channel 1 embedding, respectively. The red boxes denotes the masked windows. (D) Mean absolute error values of reconstructing each of the channel signals from different channels, computed on the test set. (E) t-SNE plot of the embeddings of the signals of different individuals from the ECG-ID dataset. (F) Improvement in the single-channel myocardial infarction diagnosis task on the PTB dataset.}
    \label{fig:enter-label}
\end{figure}

\section{Conclusion}

In this work, we have explored the possibility of correlating the representations of the different ECG channels to pave the path to a competent unified embedding and evaluated it on several independent test datasets. Our current system fails to reconstruct of sharp changes, which can be improved by refinement through a generative adversarial network \cite{Fei2023MaskedBeyond}. We believe our mode-reduction approach has the potential to deliver generalized, channel-agnostic ECG feature extractors. The broader impact of this approach is to improve the versatility of wearable and smartwatch-based ECG sensors towards medical-grade ECG sensors.

\clearpage

\bibliography{ref}
\bibliographystyle{iclr2021_conference}

\clearpage

\appendix

\section{Method Overview}

Our proposed method aims at correlating multiple representation learners, each handling a different modality.

\begin{algorithm}
\caption{Modally Reduced ECG Representation Learning Pipeline}\label{alg:cap}
\begin{algorithmic}

\State \textbf{Encoder} : $\mathcal{E}_i, i=1,2,3,\dots,12$
\State \textbf{Decoder} : $\mathcal{D}_i, i=1,2,3,\dots,12$
\State \textbf{Input signal} : $x[i], i=1,2,3,\dots,12$
\\
\For{i=1 to 12}
\State $h[i]=\mathcal{E}_i(x[i])$

\State $\hat{x}[i]=\mathcal{D}_i(h[i])$
\EndFor

\For{i=1 to 12}
\State $reconstruction\_loss += MSE(x[i],\hat{x}[i])$
\EndFor

\State $alignment\_loss = Triplet(x[i],x[j],x'[k])$

\end{algorithmic}
\end{algorithm}

\section{Architectural Details}

Our developed Masked AutoEncoder architecture is a 1-D variant of the original Masked AutoEncoder. Since we are working with 1D signals instead of 2D images, we have modified the patch encoder layer using 1D convolutions, discarding the overlapping nature, i.e., $stride = window\_len$. After the patches or windows are embedded, we use conventional sine-cosine position embedding to preserve the temporal information.

The patches equipped with temporal identity are then passed through the encoder network, which is a sequence of ViT layers. The encoder can be summarized as:

$$Encoder = [ViT\_Layer(embed\_dim=768,$$
$$num\_heads=8,$$
$$mlp\_ratio=4, $$
$$qkv\_bias=True,$$  $$norm\_layer=LayerNorm)]_{\times 12}$$

The decoder network is similar to the encoder network, only difference being reduced complexity in terms of depth and embedding dimension.

$$Decoder = [ViT\_Layer(embed\_dim=512,$$
$$num\_heads=16,$$
$$mlp\_ratio=4, $$
$$qkv\_bias=True,$$  $$norm\_layer=LayerNorm)]_{\times 8}$$

\section{Dataset description}


The pretraining data contains twelve-lead ECGs from various sources. The twelve leads are the standard ones, i.e., I, II, III, aVR, aVL, aVF, V1, V2, V3, V4, V5, V6. These signals also contain information related to cardiac abnormalities, but they were ignored in the pretraining.

The sources of the pretraining database are:

\begin{itemize}
    \item CPSC Database and CPSC-Extra Database
    \item INCART Database
    \item PTB and PTB-XL Database
    \item The Georgia 12-lead ECG Challenge (G12EC) Database
    \item Chapman-Shaoxing and Ningbo Database
    \item The University of Michigan (UMich) Database
    \item Augmented Undisclosed Database
\end{itemize}

These databases contain more than 100,000 signals, and out of them 88,000 signals are considered as the public training data, which is the split we also used for pretraining. These signals comprise individuals from four countries across three continents. 

We used the signals with sampling frequency of 500 Hz for our pretraining. INCART and PTB consists of signals of different sampling frequency which were used for testing.

\end{document}